\documentclass[fleqn,12pt,twoside]{article}
\usepackage{espcrc1}

\usepackage{graphicx}

\newcommand{\AmS}{{\protect\the\textfont2
  A\kern-.1667em\lower.5ex\hbox{M}\kern-.125emS}}

\title{Overview of PHENIX Results from the First RHIC Run \\
\smallskip
{\normalsize W.A.~Zajc,$^{8}$ for the PHENIX Collaboration:}}

\begin{document}

\maketitle


{
\scriptsize
K.~Adcox,$^{40}$
S.{\,}S.~Adler,$^{3}$
N.{\,}N.~Ajitanand,$^{27}$
Y.~Akiba,$^{14}$
J.~Alexander,$^{27}$
W.~Allen,$^{29}$
G.~Alley,$^{29}$
L.~Aphecetche,$^{34}$
% ...
\\(PHENIX Collaboration)\\
}

{\scriptsize
% ...
$^{3}$Brookhaven National Laboratory, Upton, NY 11973-5000, USA\\
% ...
$^{14}$KEK, High Energy Accelerator Research Organization, Tsukuba-shi, Ibaraki-ken 305-0801, Japan\\
% ...
$^{27}$Chemistry Department, State University of New York - Stony Brook, Stony Brook, NY 11794, USA\\
% ...
$^{29}$Oak Ridge National Laboratory, Oak Ridge, TN 37831, USA\\
% ...
$^{34}$SUBATECH (Ecole des Mines de Nantes, IN2P3/CNRS, Universite de Nantes) BP 20722 - 44307, Nantes, France\\
% ...
$^{40}$Vanderbilt University, Nashville, TN 37235, USA\\
% ...
}

\begin{abstract}
Results from the PHENIX experiment for the 
first RHIC run with Au-Au collisions at 
$\sqrt{s_{NN}} = 130$~GeV are presented. 
The systematic variation with centrality of
charged particle multiplicity, 
transverse energy, elliptic flow, identified particle spectra and 
yield ratios, 
and production of charged particles and $\pi^0$'s
at high transverse momenta are presented.
Results on two-pion correlations
and electron spectra are also provided, along with 
a discussion of plans for the second run at RHIC.
\end{abstract}

\section{INTRODUCTION}
The PHENIX experiment\cite{Morrison:1998qu} has been designed to measure a broad variety of
signals from both heavy ion and polarized proton-proton collisions at RHIC.
The pursuit of penetrating 
probes generated in the early stages of the collision, combined with
a program of hadron measurements, provides a detector with
unparalleled capabilities to address observables sensitive
to all stages of the collision process. 
This same detector is also very well-suited to the study of 
gluon and anti-quark contributions to the proton spin\cite{Saito:1998cx,Bunce:2000uv}.

The PHENIX detector consists of three spectrometers: two muon spectrometers
covering the full azimuth for  $1.1 < |\eta| < 2.4 $ and a central spectrometer
consisting of two arms each subtending $90^o$ in azimuth and with $ |\eta| < 0.35$.  
A central magnet provides an axial field, while each muon spectrometer contains
a magnet that produces a roughly radial field.
The central arms contain three tracking sub-systems: pad chambers (PC), 
drift chambers (DC)
and time-expansion chambers (TEC); two forms of electromagnetic
calorimetry (PbSc and PbGl); a time-of-flight hodoscope (TOF) and 
ring imaging Cerenkov counter (RICH).
These sub-systems, together with a set of beam-beam counters (BBC) located in 
the region $ 3 < |\eta| < 3.9$,  provide superb hadron and electron identification
over a broad range of transverse momentum\cite{Hamagaki}.
The muon spectrometers use cathode strip chambers in three
stations for tracking (muTr), and five layers of Iarocci tubes
interleaved with iron absorber for muon identification (muID).
Global event characterization is achieved via a multiplicity and vertex detector (MVD)
consisting of silicon strips and pads covering $|\eta| < 2.5$,
and the RHIC-standard Zero-Degree Calorimeters (ZDCs), which detect
neutral particles emitted along the beam directions\cite{White,Adler:2000bd}.
The front end electronics for all sub-systems are clocked synchronously
with the beam crossing frequency of 9.4 MHz.
A set of Level-1 triggers derived from various sub-systems is used to initiate
readout of the entire detector through a pipelined high bandwidth data acquisition system
capable of archiving 20 MB/s.  

\begin{figure}[htb]
 \centerline{
   \resizebox{0.445\textwidth}{!}{\includegraphics{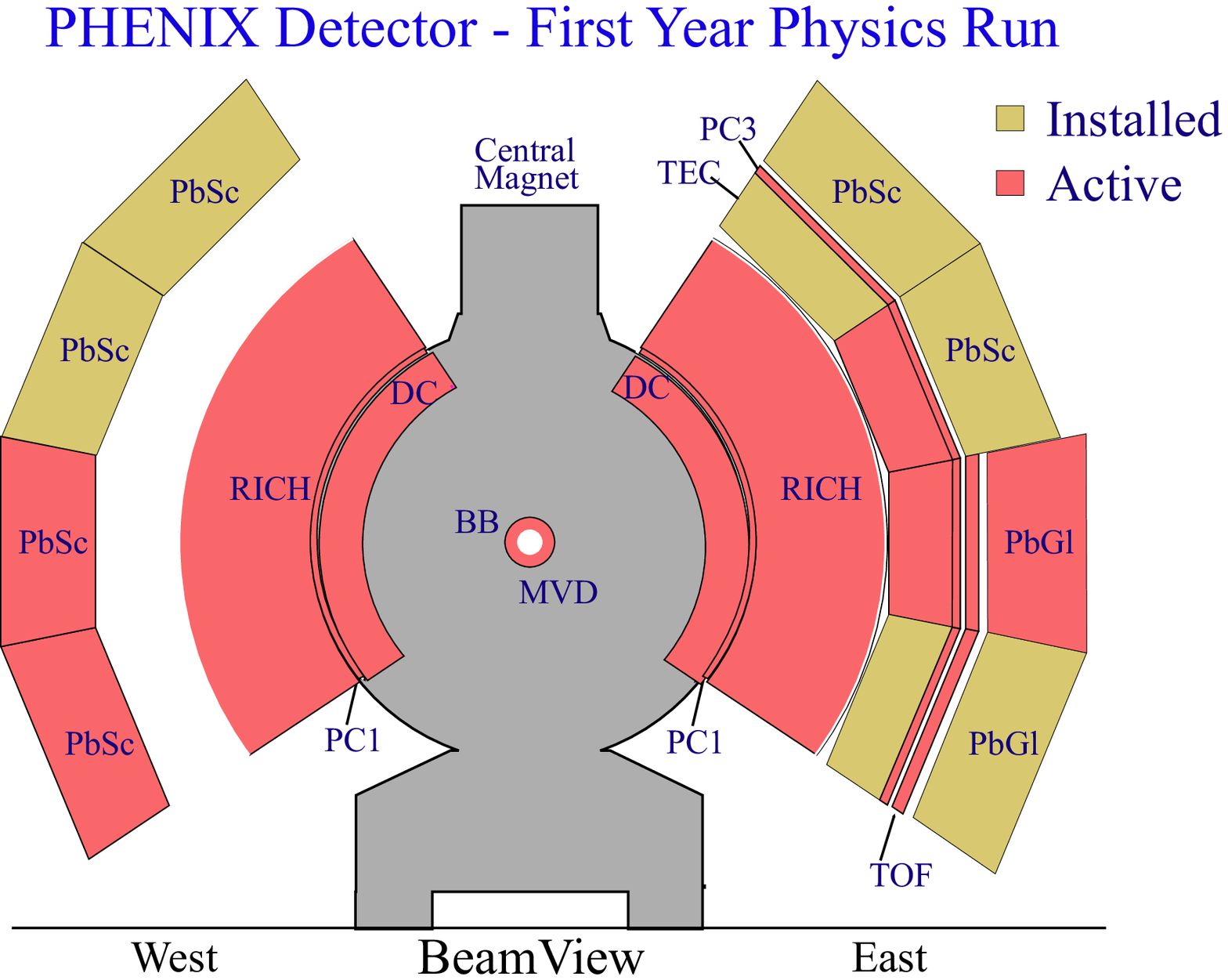}}
   \resizebox{0.555\textwidth}{!}{\includegraphics{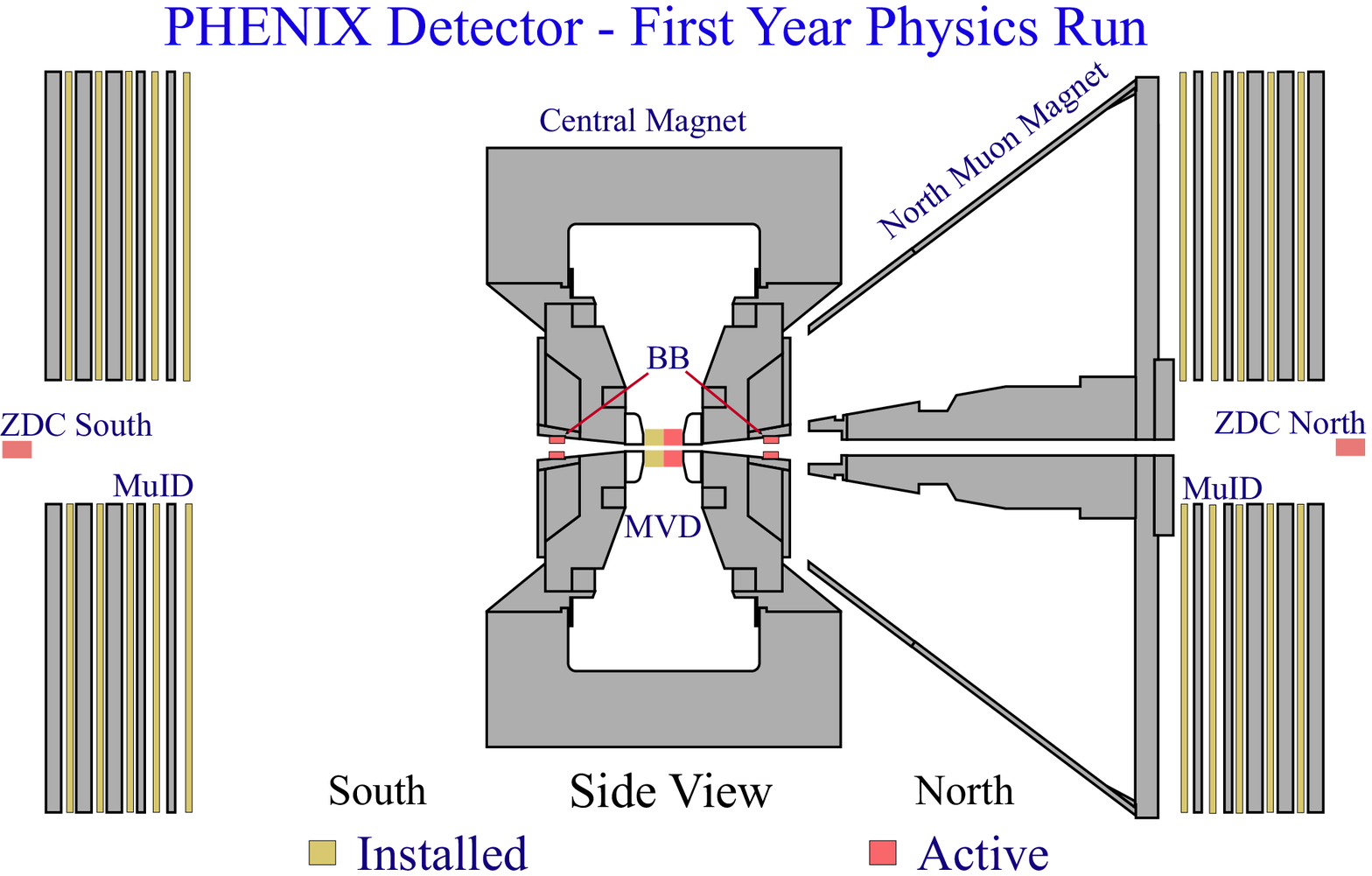}}
 }
  \caption{Installed and active detectors for  
           the RHIC Run-1 configuration of the PHENIX experiment.}
  \label{Fig:Run1}
\end{figure}
For the first physics run of RHIC in the summer of 2000, the
portions of the PHENIX detector shown in Figure~\ref{Fig:Run1} were instrumented. 
Elements of all sub-systems, with the exception of the muTr, were in place and read out. 
Small subsets of the MVD and muID front end electronics were connected
and tested as part of an engineering run. 
All other sub-systems were instrumented in fractions ranging from 25\%
to 100\% of their ultimate aperture and were used in the physics results which are
presented here.
Independent minimum bias triggers were formed using coincidences between the BBC counters
and between the ZDC counters. 
A total of approximately 5M events  was recorded  
at $\sqrt{s_{NN}}=130$~GeV.
The primary trigger used for most of the results
presented below is based on the BBC coincidence with an additional
offline requirement 
that restricts
the collision vertex to $|z| < 20$~cm. 
Whenever possible, physics quantities are presented as a function
of centrality and/or the number of participants $N_{part}$.
Details on the determination of these quantities are presented in 
the following section and in the associated references.

\section{GLOBAL OBSERVABLES}
\label{Sec:Global}

The systematic variation of 
particle yields and the produced transverse energy with the number
of participants reflects the underlying reaction mechanisms.
For example, Gyulassy and Wang\cite{Wang:2001bf} have emphasized that such a study
can discriminate between cascade models and models which incorporate
gluon saturation effects. 
PHENIX has studied the production near mid-rapidity of both charged particles
and of transverse energy as a function of centrality.  
The deposition of energy in the ZDC's is correlated with that of
charge in the BBC's to provide an unambiguous mapping between
these observables and the centrality of the collision,
\begin{figure}[htb]
\begin{minipage}[b]{0.6\linewidth}
  \mbox{\resizebox{\textwidth}{!}{\includegraphics{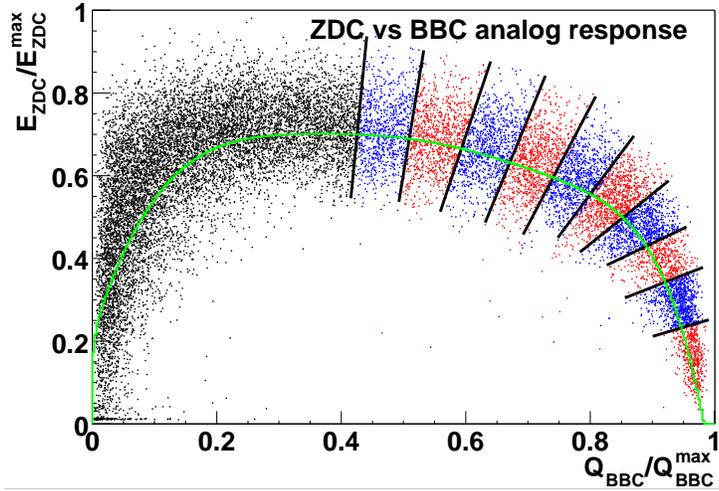}}}
\end{minipage}
\hfill
\parbox[b]
 {.3\textwidth}{\sloppy
  \caption{BBC vs ZDC analog response, related to the centrality classes
           of the collisions. The rightmost interval is the  \hbox{0-5\%} 
           centrality class,
           the next is \hbox{5-10\%} and so on.}
  \protect\label{Fig:porcupine}
 }
\end{figure}
as shown in Figure~\ref{Fig:porcupine}.
A Glauber model is then used to determine the number of participants
$N_P$ and the number of binary collisions $N_C$ for each centrality class. 
(This methodology is used consistently for all such studies presented
in this contribution.)

\begin{figure}[htb]
 \centerline{
   \resizebox{0.5\textwidth}{!}{\includegraphics{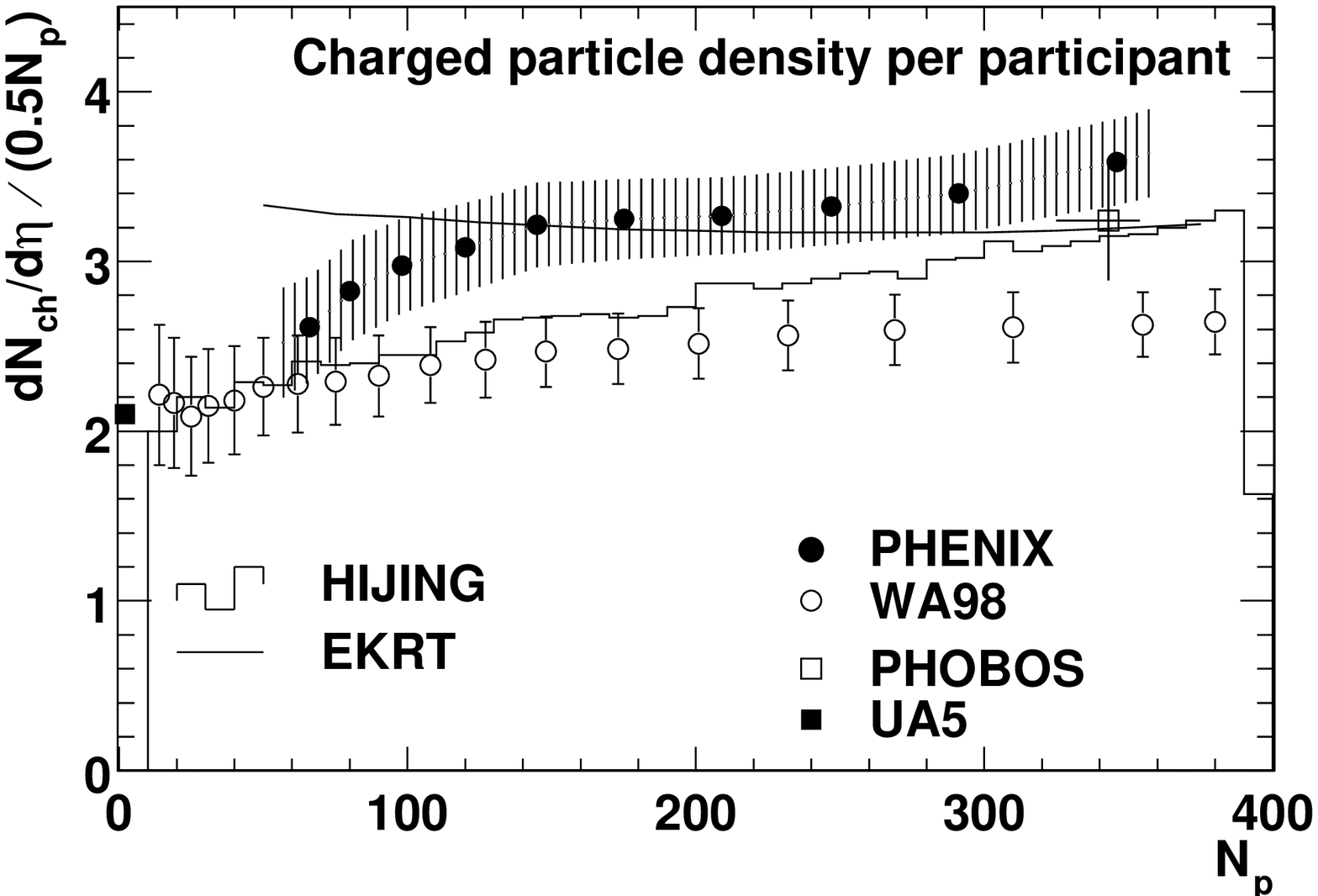}}
   \resizebox{0.5\textwidth}{!}{\includegraphics{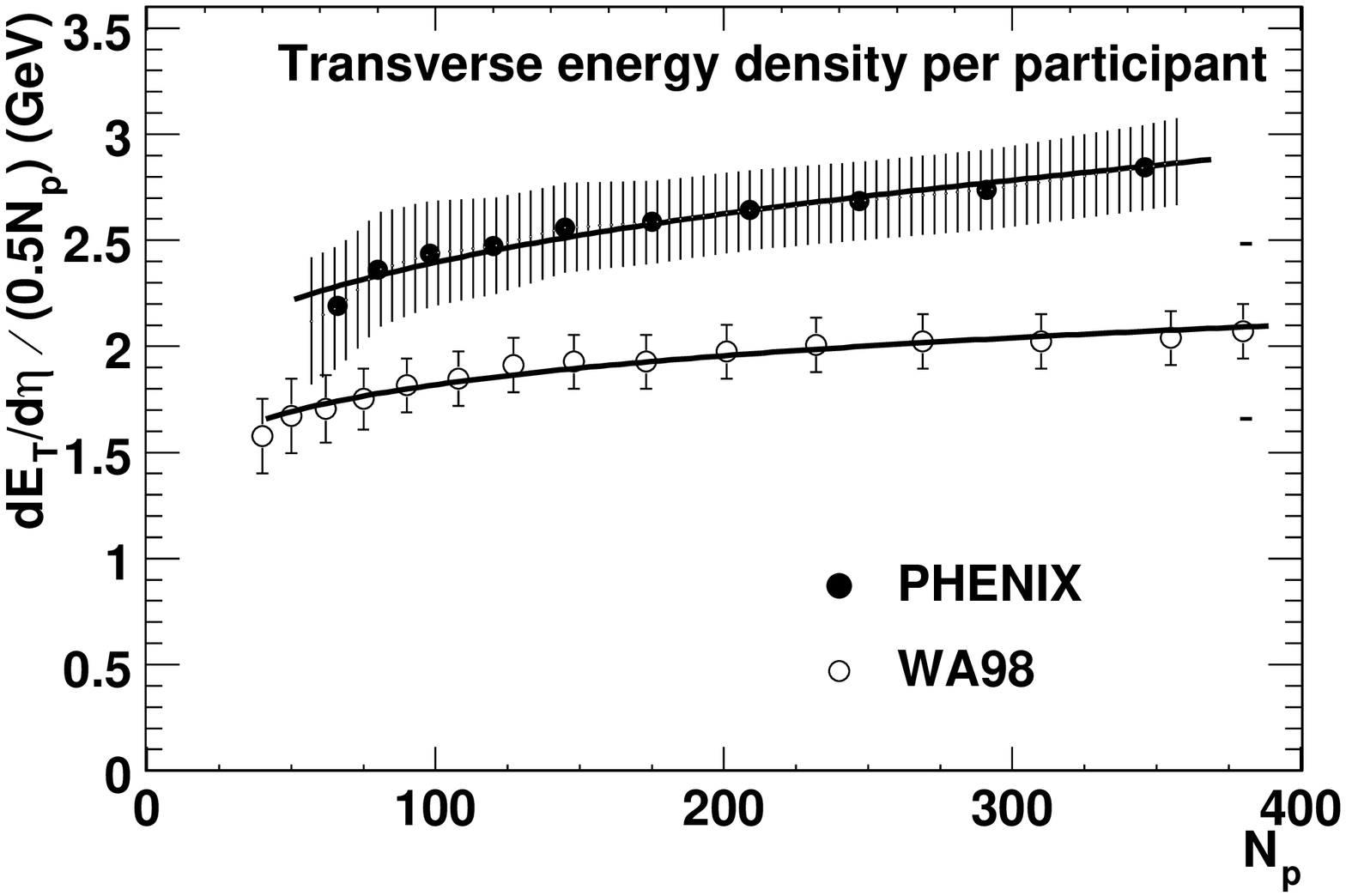}}
 }
 \caption{a.) The pseudorapidity density per participant pair versus the number of
 participants. b.) The transverse energy pseudorapidity density per participant pair versus
 the number of participants. The band in each figure indicates the systematic errors.} 
 \label{Fig:StuffVsNpart}
\end{figure}
The number of produced charged particles is determined from 
the correlation between hits in two layers of Pad Chambers with the 
vertex location, which provides (on a statistical basis)
the charged particle multiplicity distribution in the interval
$|\eta| < 0.35$. 
After corrections for acceptance, efficiencies,
decays and double hits, the charged particle pseudo-rapidity 
distribution $dN_{ch}/d\eta$ is calculated for each centrality
bin, scaled by the corresponding number of participant pairs, then plotted 
versus $N_P$, as shown in Figure~\ref{Fig:StuffVsNpart}a.
Also shown there are comparisons to the model predictions of
EKRT\cite{Eskola:2001xq,Eskola:2000fc}, 
which does not reproduce the trend of the data.
(Results very similar to the PHENIX data were reported
at this conference by the PHOBOS Collaboration\cite{PHOBOS}.)

A similar analysis has been performed for transverse energy measured
in the PHENIX PbSc calorimeter\cite{Adcox:2001ry}. 
A careful treatment of the contributions from produced energy
in the aperture, the in-flux from scattering sources, and both
the in-flux and out-flux from decays is performed to convert
the deposited energy seen in the calorimeter to the equivalent
transverse energy. The total transverse energy density per participant pair
is then calculated in the same $N_P$ bins as used in the multiplicity
analysis. The trend,  shown in Figure~\ref{Fig:StuffVsNpart}b., 
is essentially identical to that found for charged multiplicity,
and indicates that particle and transverse energy production is not
simply proportional to the number of participants.
The yield may be described as a superposition of terms proportional
to participants and binary collisions, 
\begin{equation}
 \quad \quad
 {\rm Yield } = A \cdot N_P + B \cdot N_C \quad.
 \label{Eq:AB}
\end{equation} 
The results of this procedure are given in Table~\ref{Tab:AB}.
The extent to which the ratio $B/A$ is significantly different from zero 
may be taken as evidence for the role of binary collisions in contributing
to production of particles and transverse energy at RHIC\cite{Milov}. 
\begin{table}[htb]
\caption{}
\newcommand{\cc}[1]{\multicolumn{1}{c}{#1}}
\renewcommand{\tabcolsep}{2pc} 
\renewcommand{\arraystretch}{1.2} 
\begin{tabular}{llll}
\hline
Quantity            &       \cc{$A$}         &      \cc{$B$}      &       \cc{$B/A$} \\
\hline
$  dN_{ch} / d\eta  $  & $0.88 \pm 0.28$     & $0.34\mp 0.12$     &  $0.38\pm 0.19$  \\
$  dE_{T}  / d\eta  $  & $0.80 \pm 0.24$ GeV & $0.23\mp 0.09$ GeV &  $0.29\pm 0.18$  \\
\hline
\end{tabular}\\[6pt]
The results of fits to Equation~\ref{Eq:AB} for charged multiplicity
and transverse energy near $\eta=0$.
Note the dimensions of $A$ and $B$ are different for the two analyses. The $\mp$ symbol in the error
for $B$ is used to indicate that its error is largely anti-correlated with that of $A$. 
This anti-correlation is propagated in calculating the error in the ratio $B/A$.
\label{Tab:AB}
\end{table}

Also shown in Figure~\ref{Fig:StuffVsNpart} are comparisons of both the charged
multiplicity and the transverse energy per participant pair to the
distributions measured by the WA98 collaboration at the CERN SPS\cite{WA98},
clearly demonstrating the increase of both quantities at RHIC. In the case
of energy density calculated following the Bjorken prescription\cite{Bj83},
the value of 4.6~${\rm GeV/fm^3}$ is roughly 60\% larger\cite{Adcox:2001ry} 
than that found at the SPS\cite{NA49ET}. It is interesting to note that this increase does
not result from an increase in 
$\langle dE_T/d\eta \rangle / \langle dN_{ch}/d\eta \rangle$ ;
the value is very nearly equal to that found at the SPS\cite{Adcox:2001ry}.

\begin{figure}[htb]
  \resizebox{\textwidth}{!}{\includegraphics{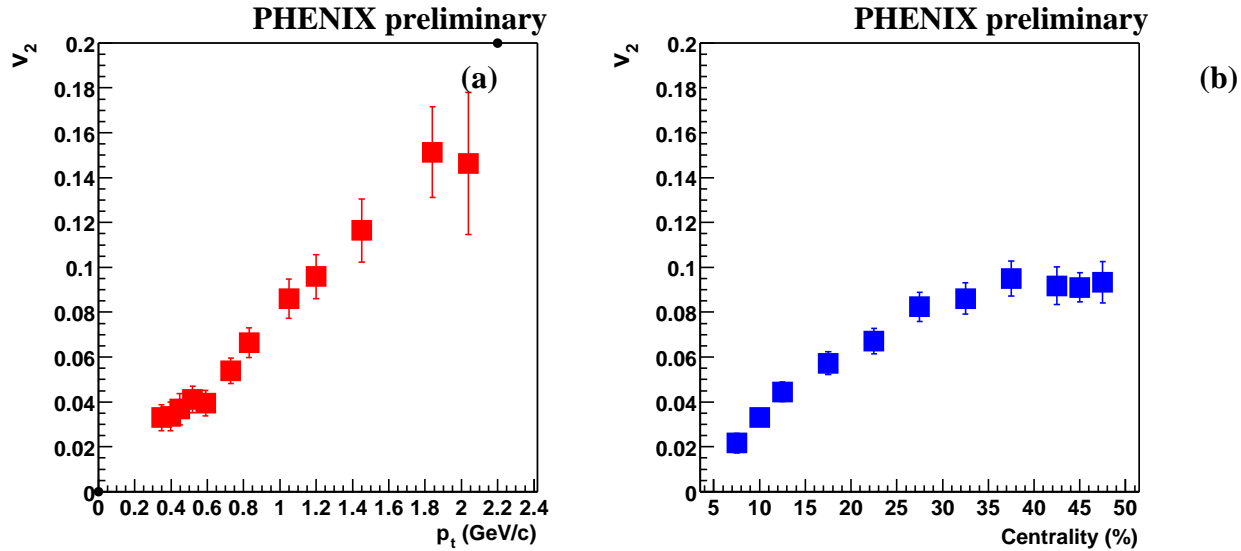}}
  \caption{$v_2$ versus transverse momentum (a) and centrality (b).
           Errors are statistical only.}
  \label{Fig:Flow}
\end{figure}
Another global feature of hadronic production is the azimuthal pattern of
emission, as parameterized by elliptic flow. Typically such analyses are 
performed with respect to the reaction plane determined for each event.
It is also possible to determine the elliptic flow pattern by measuring
the auto-correlation function of particles in the event, that is, by
calculating the correlation function $C(\Delta\phi)$, where $\Delta\phi$
is the difference in azimuthal angle between two particles from the same
event. PHENIX has performed such an analysis using charged particles
with $p_T > 200$~MeV/c in the interval $|\eta| < 0.35$. The correlation function is calculated
using an event-mixing prescription to determine the background
distribution, then fit to extract the (assumed positive) $v_2$ coefficient:
\begin{equation}
\quad \quad 
C(\Delta\phi) \propto 1 + 2 |v_1|^2 \cos(\Delta\phi) + 2 |v_2|^2 \cos(2\Delta\phi) 
\end{equation}
Standard PHENIX event characterization methods are used to study the dependence
of $v_2$ versus centrality in each $p_T$ bin. A sample of the results\cite{Lacey} are shown
in Figure~\ref{Fig:Flow}. These trends, which are in good agreement with data
from STAR\cite{Ackermann:2001tr} and preliminary results from PHOBOS\cite{PHOBOS}, 
suggest that the high density matter formed at RHIC efficiently translates the
initial spatial asymmetry into a corresponding one in momentum space.

\section{IDENTIFIED HADRONS}
\begin{figure}[htb]
  \resizebox{\textwidth}{!}{\includegraphics{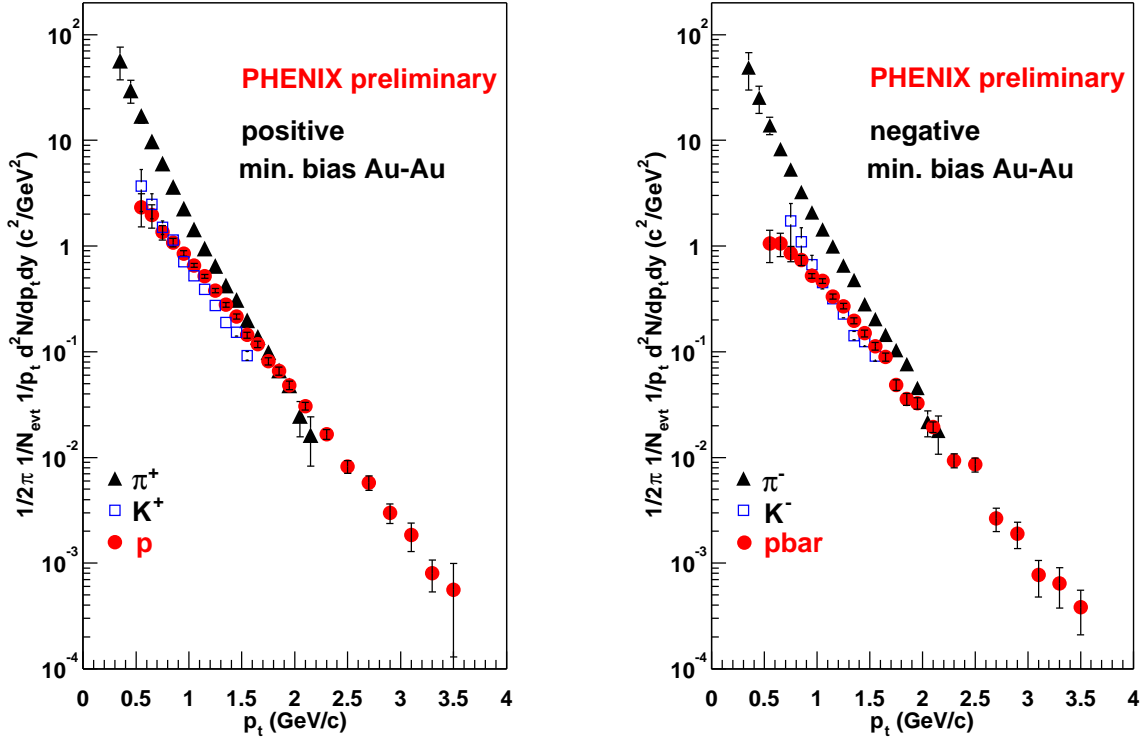}}
  \caption{Normalized minimum bias transverse momentum spectra for positive (left) and negative 
           (right) identified particles. The error bars are composed of statistical errors
           and the systematic errors associated with acceptance and decay corrections.
           There is an additional 20\% systematic error associated with the overall normalization.}
  \label{Fig:SPECTRAFig1.ps}
\end{figure}
The PHENIX central arm detectors have been designed to provide particle identification
over the broadest possible momentum range\cite{Hamagaki}. 
The primary tool for charged hadron identification is the time-of-flight
difference between the BBC and the highly-segmented TOF hodoscope, which spans
$\Delta\phi = 45^o$ in the East spectrometer arm. 
The overall time resolution $\sigma \sim 115$~ps permits unambiguous $\pi/K$ separation
to at least $p_T = 1.5$~GeV/c.
Normalized minimum bias $p_T$ spectra for 
$\pi^\pm$, ${\rm K}^\pm$, $p$'s and ${\bar p}$'s are shown in Figure~\ref{Fig:SPECTRAFig1.ps}.
The shape of the spectra clearly depends on the particle species, with pions having
the lowest $\langle p_T \rangle$ and protons and anti-protons having the largest.
The dependence of the local slopes increases with both centrality and with particle
mass, consistent with expectations from radial flow\cite{Velkovska}.

\begin{figure}[htb]
 \centerline{
   \resizebox{0.50\textwidth}{!}{\includegraphics{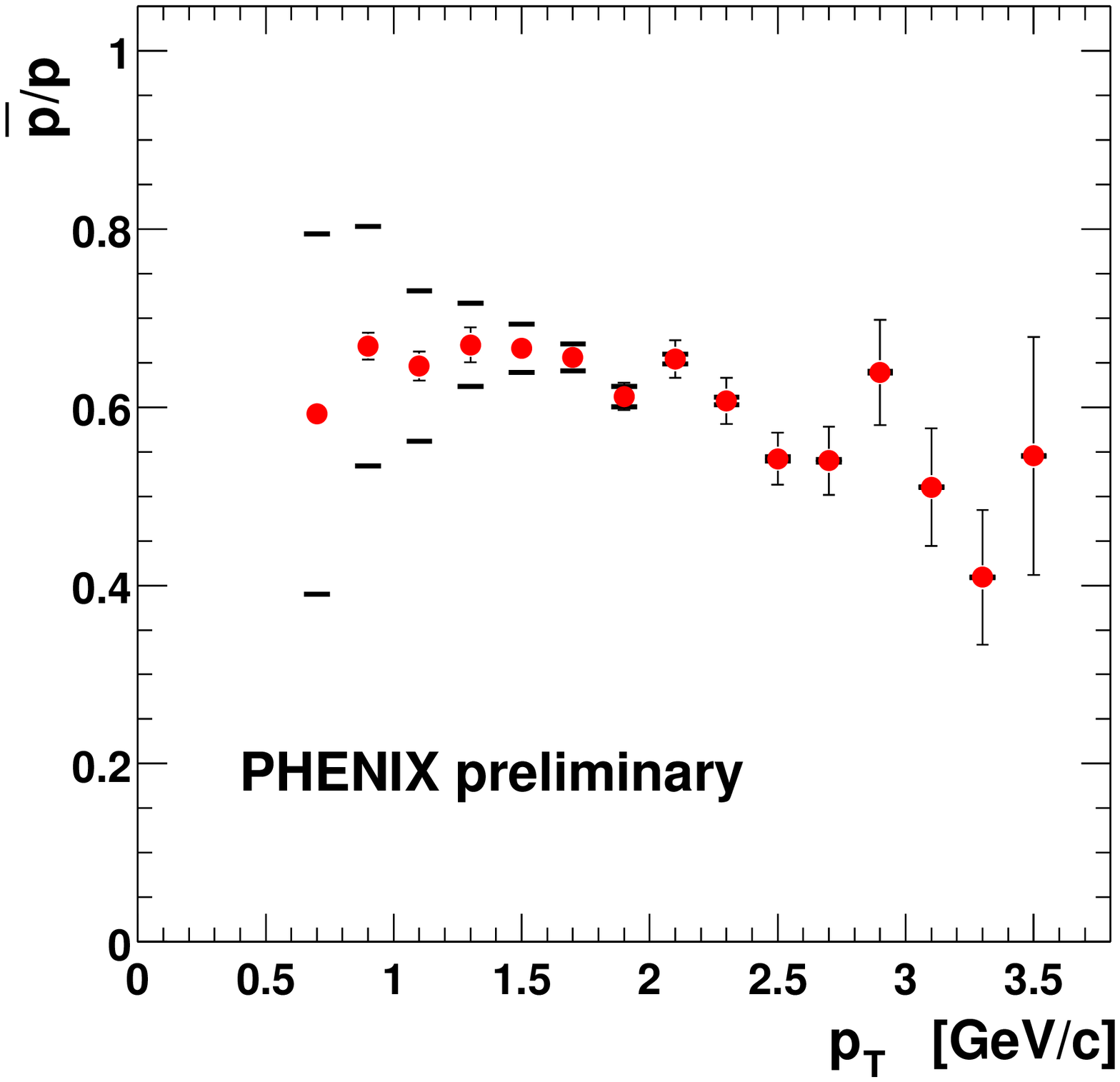}}
   \resizebox{0.50\textwidth}{!}{\includegraphics{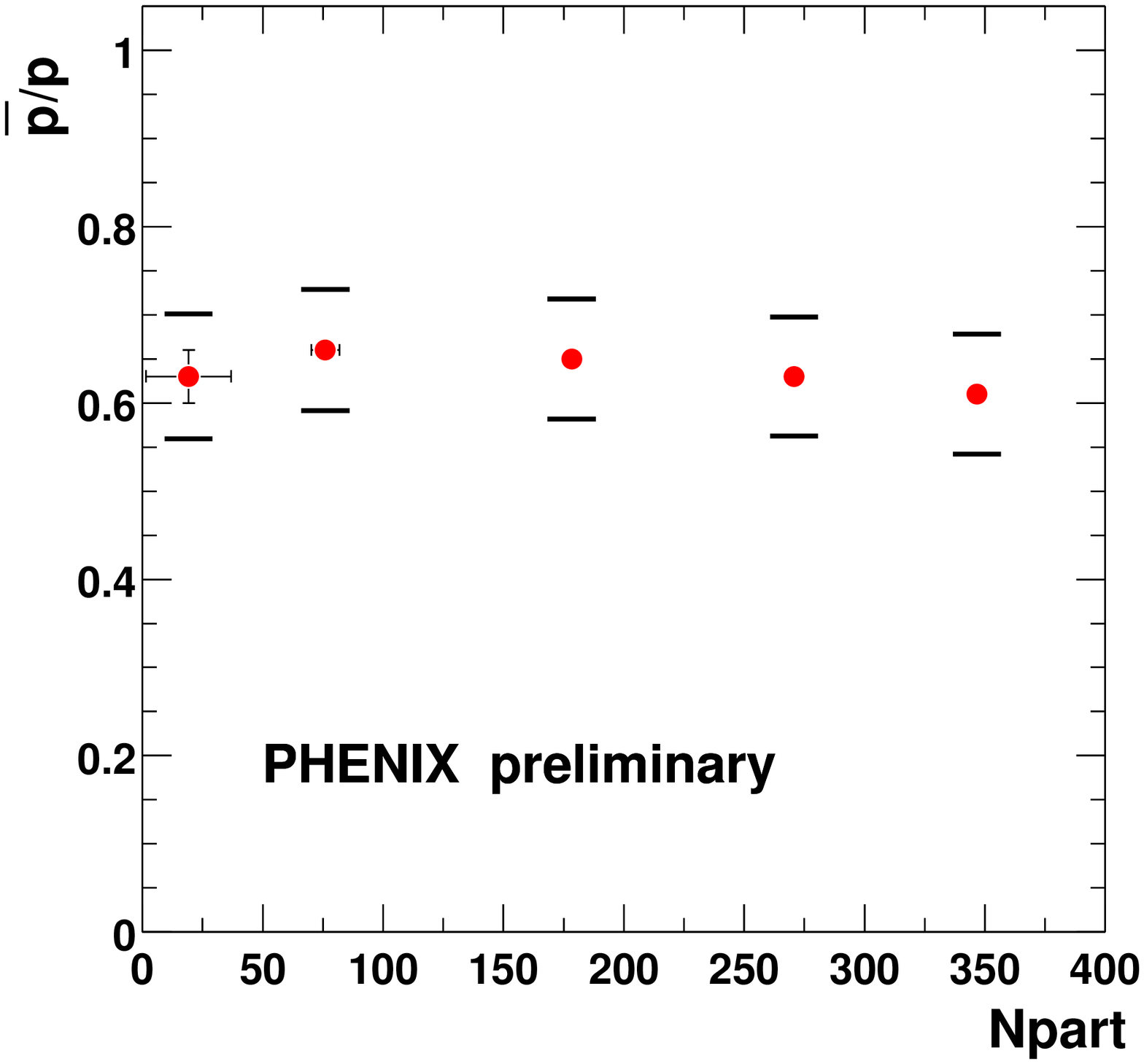}}
 }
  \caption{The ${\bar p}/p$ ratio versus transverse momentum for minimum bias
           Au-Au collisions (left) and versus centrality (right).
           For both figures, the error bars represent that statistical error on the ratio;
           systematic errors are indicated by the horizontal bars.}
  \label{Fig:pbarp}
\end{figure}
The ratio of anti-protons to protons at $y=0$ is of particular interest, since it 
is a direct measure of the net baryon content in the central region. 
The ${\bar p}/p$ ratio has been studied as a function of both transverse momentum
and centrality\cite{Ohnishi}, and found to be only weakly dependent on $p_T$ and independent 
of centrality within systematic
errors (Figure~\ref{Fig:pbarp}).
For minimum bias collisions, the ${\bar p}/p$ ratio 
in the interval 0.8 GeV/c $< p_T <$ 3.0 GeV/c is  0.64 $\pm$ 0.01(stat.) $\pm$ 0.07(sys.).
This value 
(which is consistent with that observed by the 
other RHIC experiments\cite{PHOBOS,STARpbar,BRAHMSpbar})
together with the measured yields shown in Figure~\ref{Fig:SPECTRAFig1.ps} 
signifies that the central region in heavy ion collisions at RHIC is
meson-dominated, as distinct from the baryon-dominated case
at the SPS and lower energies, where the ratio is never
greater than $\sim 0.1$\cite{NA44pbar,NA49pbar}.

\begin{figure}[htb]
\begin{minipage}[b]{0.55\textwidth}
  \mbox{\resizebox{1.2\textwidth}{!}{\includegraphics{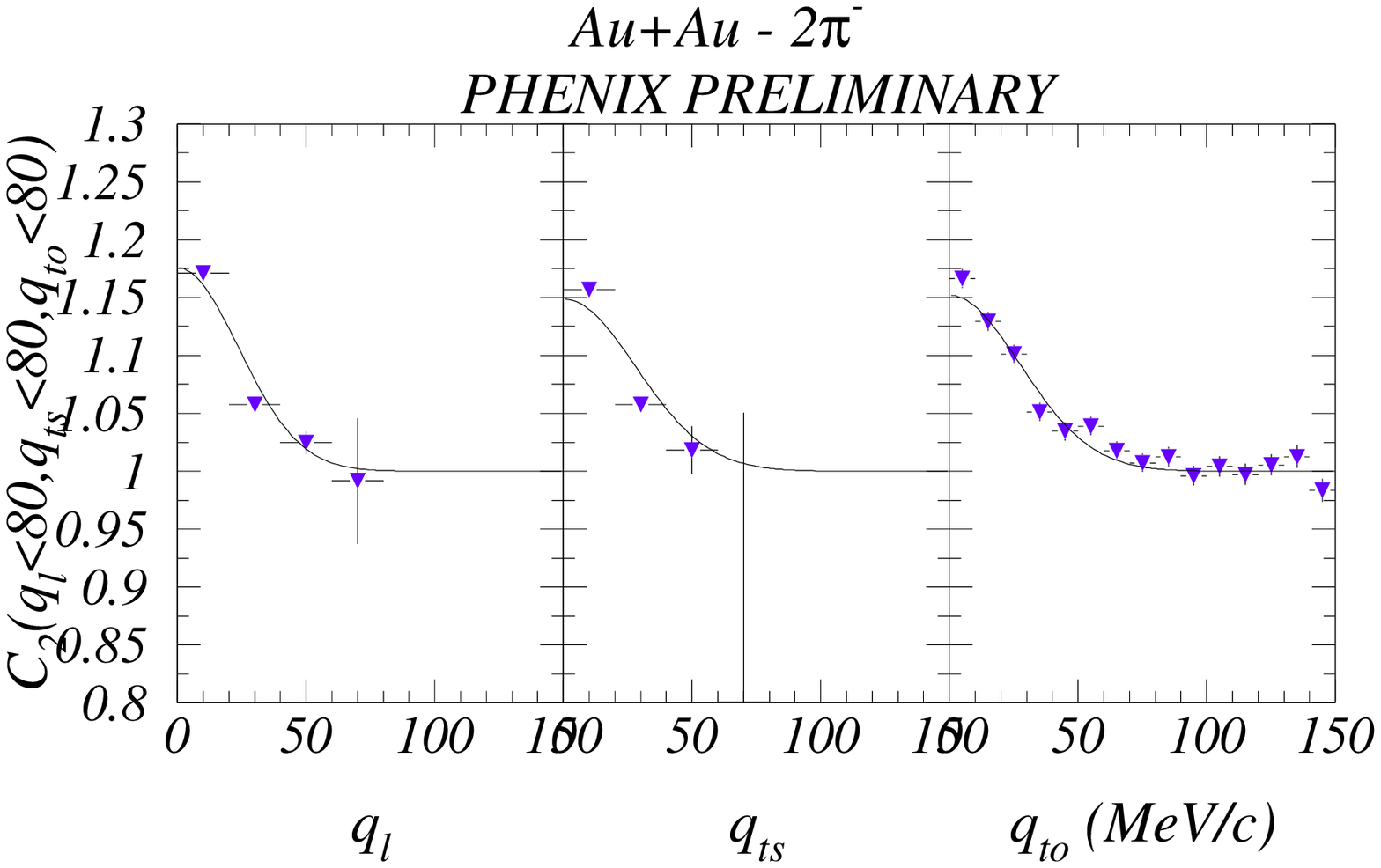}}}
\end{minipage}
\hfill
\parbox[b]
 {.35\textwidth}{\sloppy
  \caption{Correlation function for $\pi^-$ pairs from the EMC
analysis.  The one-dimensional projections of the three-dimensional
correlation function are averaged over the lowest 80 MeV/c in the other
momentum differences. The fits are performed over the full phase space;
the curves in the projections are averaged in the same manner as the data.}
  \protect\label{Fig:HBT}
 }
\end{figure}

\begin{table}[htb]
\caption{}
\newcommand{\cc}[1]{\multicolumn{1}{c}{#1}}
\renewcommand{\tabcolsep}{1.5pc} 
\renewcommand{\arraystretch}{1.2} 
\begin{tabular}{c|cccc}\hline
Data Set & $R_{Tout}$ (fm) & $R_{Tside}$ (fm) & $R_{Long}$ (fm) & $\lambda$ \\ \hline
\hline
EMC $\pi^+\pi^+$ & $4.4 \pm 0.2$ & $5.1 \pm 0.6$ & $5.9 \pm 0.4$ & $0.27
\pm .02$ \\ \hline
TOF $\pi^+\pi^+$ & $6.2 \pm 0.5$ & $7.9 \pm 1.1$ & $4.0 \pm 1.2$ & $0.49
\pm .05$ \\ \hline
EMC $\pi^-\pi^-$ & $5.1 \pm 0.2$ & $5.0 \pm 0.6$ & $5.9 \pm 0.4$ & $0.30
\pm .02$ \\ \hline
TOF $\pi^-\pi^-$ & $5.5 \pm 0.5$ & $5.8 \pm 1.5$ & $6.7 \pm 0.9$ & $0.49
\pm .06$ \\ \hline
\hline
\end{tabular}\\[6pt]
Results of the Bertsch-Pratt fits to the identical pion pairs
in the EMC and TOF analyses.  
Only statistical errors are shown; current systematic uncertainties are $<1$ fm.
\label{Tab:HBT}
\end{table}
Charged pion pairs  in the central arms have also been used to perform an
HBT analysis. Two separate analyses were performed, one using pions identified
with the TOF as described above, and the second in which the 700~ps time-of-flight
resolution of the PbSc EmCal (EMC) is used to identify pions with
transverse momentum  $p_T < 0.7$~GeV/c.  
In addition to the obvious utility of comparing two independent data sets for this analysis, 
the larger acceptance of the EMC substantially increases the number of available pairs
(by a factor of 5).
Correlation functions are calculated in the standard
$q_{Tside}, q_{TOut}, q_{Long}$ projections of the relative momentum
for both
$\pi^+\pi^+$ and $\pi^-\pi^-$ pairs and are then fit over the full 3D phase space
to the form
\begin{equation}
C_2(q_{Tside}, q_{TOut}, q_{Long}) = 
1 + \lambda \exp[ -(q_{Tside}R_{Side})^2  
                   -(q_{TOut} R_{Out} )^2
                   -(q_{Long} R_{Long})^2 
                ]
\quad .
\end{equation} 
The results, presented in Figure~\ref{Fig:HBT} and Table~\ref{Tab:HBT}, 
show little if any variation from values obtained at lower energies.
Given the significantly higher multiplicities and densities
observed at RHIC, 
this is somewhat puzzling, and perhaps indicates
significant dynamic effects on the radii from 
strong transverse expansion driven by these higher densities.
In this context, it should be noted that  
the values of the radii and $\lambda$ (for the TOF analysis), 
as well as the dependence on pair
transverse momentum\cite{Johnson} are all consistent
with measurements from the STAR collaboration\cite{Harris}.

\section{PARTICLE PRODUCTION AT HIGH TRANSVERSE MOMENTUM}
\label{Sec:HighPT}
The high center-of-mass energy provided by RHIC, corresponding
to values of $\sqrt{s_{NN}}$ where hard scattering
at the partonic level is observed in $p$-$p$ and
$p$-${\bar p}$ collisions,
offers the exciting possibility of using perturbative probes
amenable to quantitative calculation to explore hot nuclear matter.
A first step in this program 
is the measurement of the transverse
momentum spectrum for charged particles and its variation with the number
of participants in the collision.
PHENIX has performed such an analysis using charged tracks reconstructed
with the drift and pad chambers of the central arm\cite{Messer}.
The raw distribution of tracks is formed for six exclusive centrality
classes, again using the standard PHENIX event classification 
described in Section~\ref{Sec:Global}.
Corrections are made in each centrality bin for the spectral
distortions due to momentum resolution and to backgrounds
from scattering and decays.
\begin{figure}[htb]
\begin{minipage}[t]{0.4\textwidth}
\includegraphics[width=1.0\textwidth]{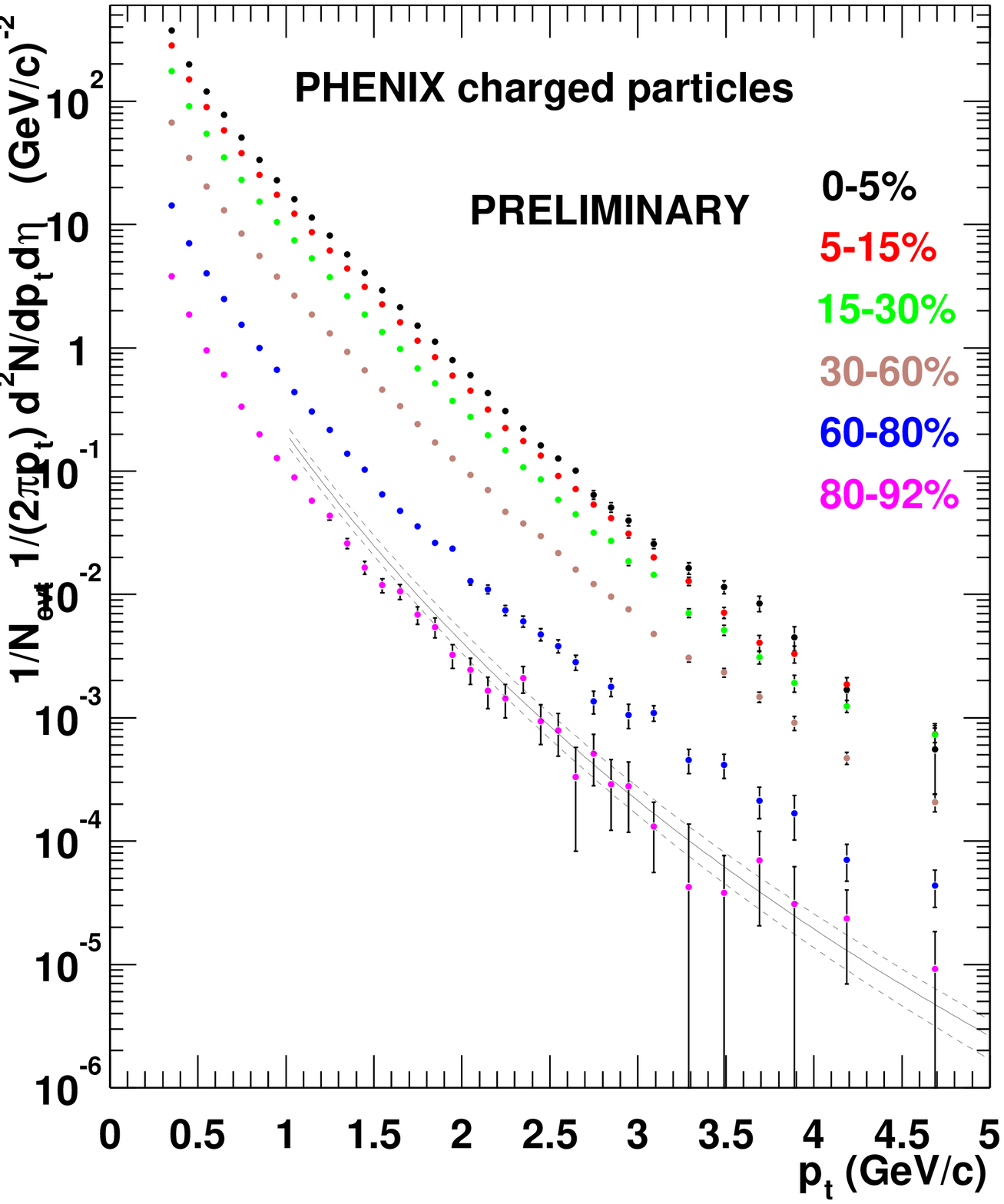}
 \caption{Normalized invariant transverse momentum spectra 
          in the interval $|\eta| < 0.35$, for six centrality
          classes. The curve compares the most peripheral class (80-92\%) to 
          the yield expected from p-p collisions scaled by the corresponding
          number of binary collisions.}
\label{Fig:ChargedSpectra}
\end{minipage}
\hspace{\fill}
\begin{minipage}[t]{0.5\textwidth}
\includegraphics[width=1.0\textwidth]{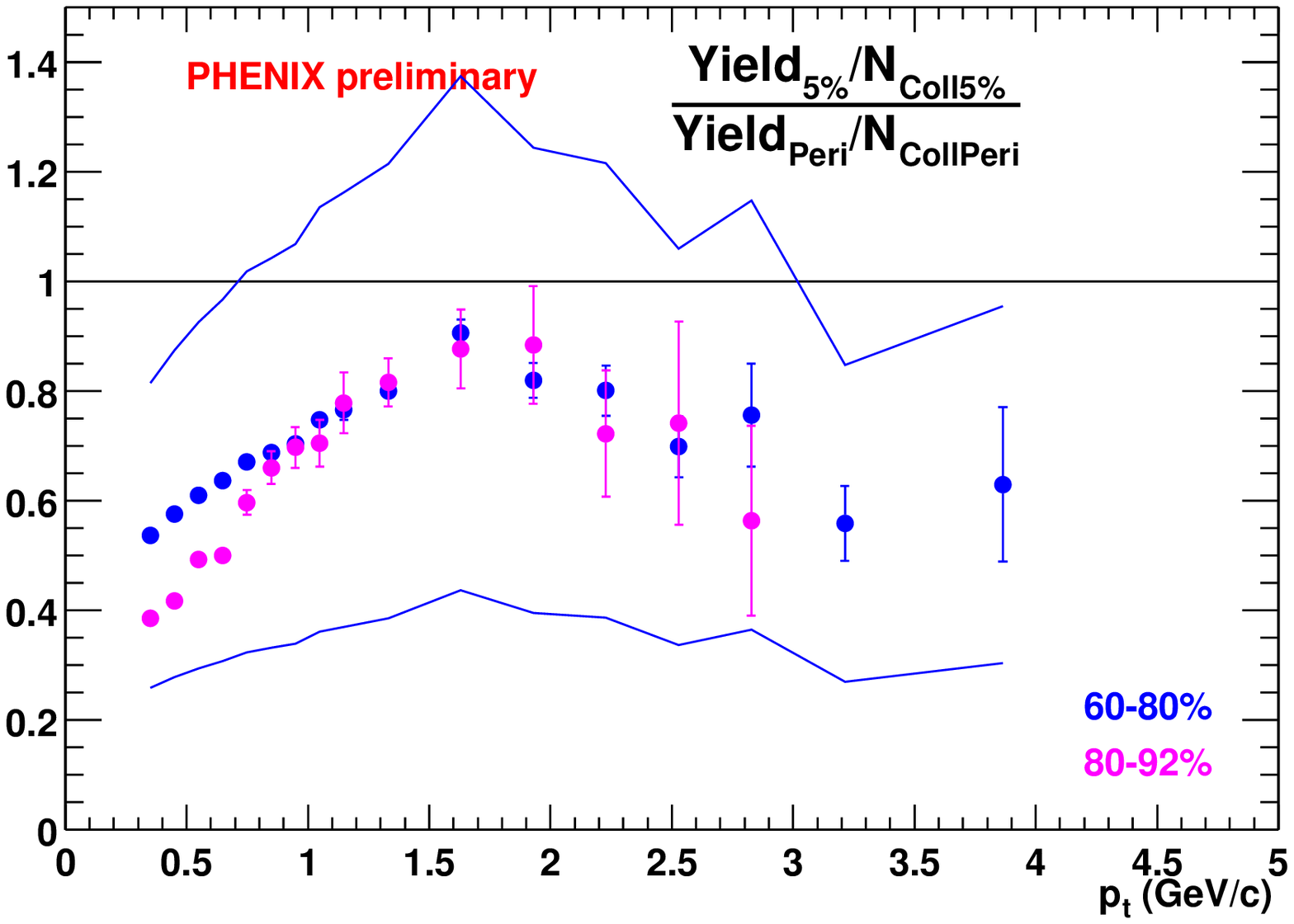}
\caption{The $p_T$ dependence of the ratio of yields for central (0-5\%) collisions
         to yields for two peripheral classes (60-80\% and 80-92\%),
         scaled by the corresponding mean number of binary collisions
         for each centrality class.}
\label{Fig:ChargedRatio}
\end{minipage}
\end{figure}
Results are shown in Figure~\ref{Fig:ChargedSpectra}
for the interval $0.4 < p_T < 5$~GeV/c, which spans six decades
in yield. 

In nucleon-nucleon collisions, production of high transverse momentum particles
results from hard scattering at the partonic level. Since these
cross sections are small, their contribution in nucleus-nucleus
collisions, absent collective effects, should scale as the
total number of binary collisions. For the most peripheral (80-92\% in centrality) bin
shown in Figure~\ref{Fig:ChargedSpectra},
the shape and yield above $p_T > 2$~GeV/c are in good agreement 
with an interpolation of hadron-hadron data\cite{Drees} scaled by the number
of binary collisions ($\langle N_{coll} \rangle = 3.7 \pm 2$)
calculated for this centrality bin. This is not the case for the 
most central collisions, which show a deficit at large $p_T$ with
respect to $N_{coll}$ scaling. This is presented in Figure~\ref{Fig:ChargedRatio},
where the scaled ratio
\begin{equation}
\quad \quad
{ 
  { Yield(Central)    \ / \ \langle N_{coll}(Central)    \rangle }
  \over
  { Yield(Peripheral) \ / \ \langle N_{coll}(Peripheral) \rangle }
}
\label{Eq:ScaledRatio}
\end{equation} 
is plotted for two different values of the peripheral reference
set.
The suppression at $p_T > 3$~GeV/c is inconsistent both with enhancements
expected in that region for the Cronin effect\cite{Cronin}
and also seen in PHENIX data for mid-peripheral
data when scaled to the p-p reference distribution\cite{Messer}.
Similar results for such a suppression pattern in unidentified
charged particles have been reported by the STAR collaboration\cite{Harris}.

An independent analysis by PHENIX provides further insight into
the nature of the suppression\cite{David}. The superb segmentation
$\Delta \eta \times \Delta \phi = 0.01 \times 0.01$
and excellent resolution
$8.2\%/\sqrt{E\ {\rm (GeV)}} \oplus 1.9\%$
of the PbSc electromagnetic calorimeter are used to extract
the $\pi^0$ $p_T$ spectrum  by reconstruction (on a statistical basis)
of their principal $\pi^0 \rightarrow \gamma\gamma$ decay mode.
Extensive studies based on the mixing of single photon showers into
both real and simulated events are used to determine
the variation of the reconstruction efficiency with multiplicity
and of the background with the event multiplicity.
\begin{figure}[htb]
\begin{minipage}[t]{80mm}
\includegraphics[width=17pc]{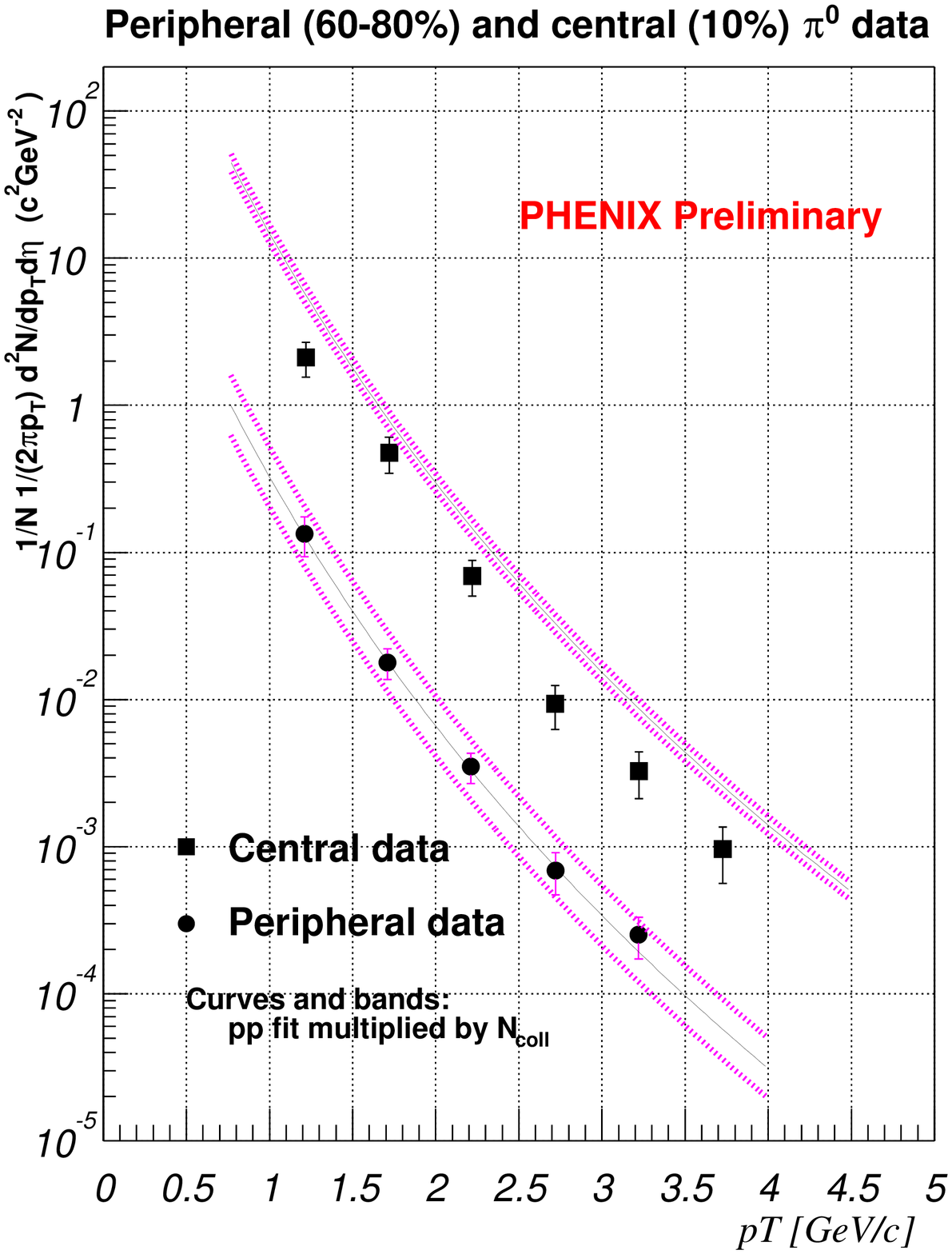}
 \caption{Normalized $p_T$ spectra for $\pi^0$'s from peripheral 
          \hbox{(60-80\%)} and central \hbox{(0-10\%)} Au-Au collisions.
          The bands compare the data to
          the yield expected from p-p collisions scaled by the corresponding
          number of binary collisions; their width indicates the systematic
          uncertainty in the number of collisions used for scaling.}
\label{Fig:Pi0Spectra}
\end{minipage}
\hspace{\fill}
\begin{minipage}[t]{75mm}
\includegraphics[width=17pc]{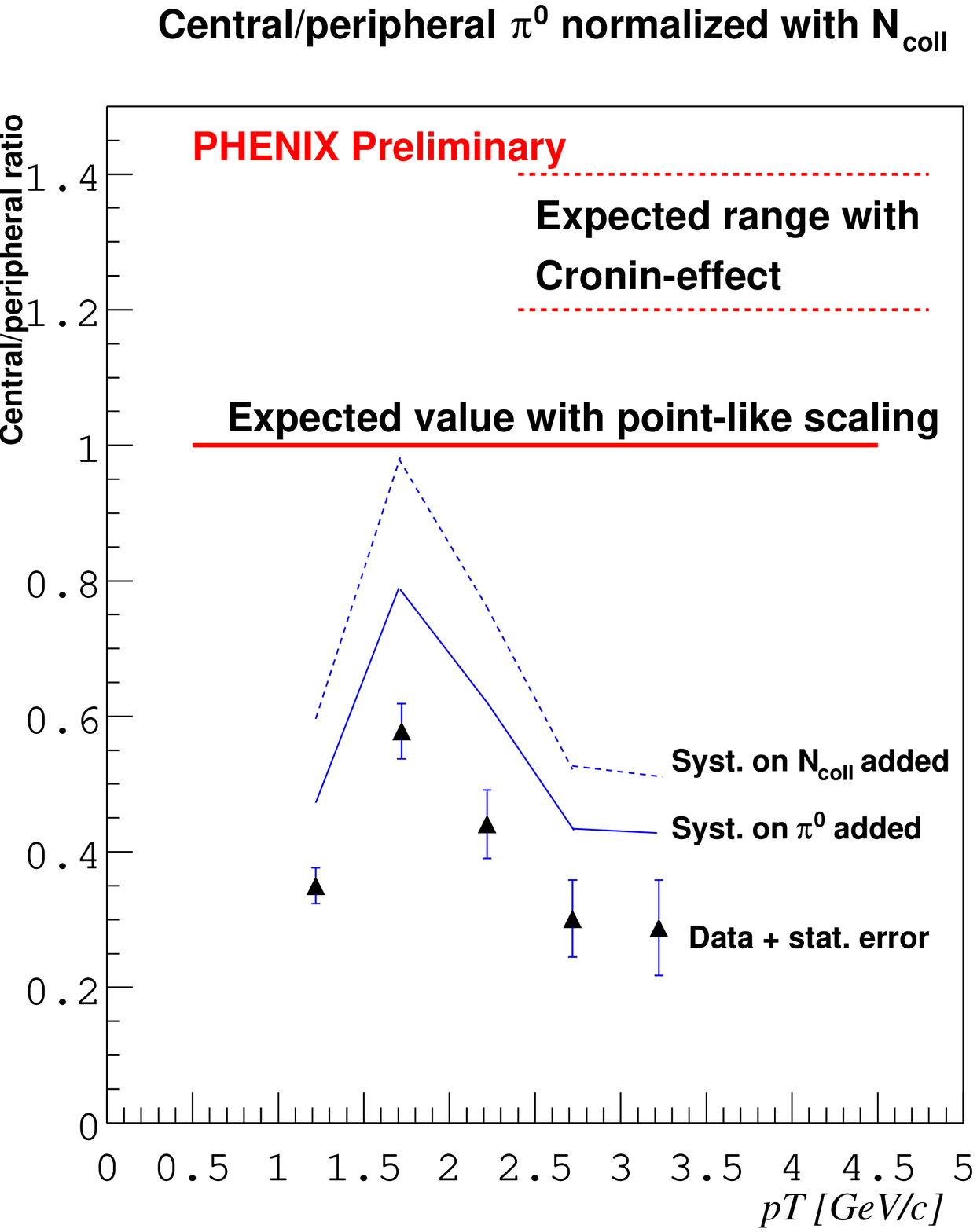}
\caption{The $p_T$ dependence of the ratio of $\pi^0$ yields for central (0-10\%) collisions
         to yields for a peripheral collisions (60-80\%),
         scaled by the corresponding mean number of binary collisions
         for each centrality class. The solid line represents the systematic error on
         the yields; the dashed line the additional systematic error from the uncertainty in
         the number of collisions used for scaling.}
\label{Fig:Pi0Ratio}
\end{minipage}
\end{figure}
A detailed simulation is used to estimate the contributions
from background and decay particles.
Results are shown in Figure~\ref{Fig:Pi0Spectra} for both a 
peripheral (60-80\%) and central (0-10\%) sample.
As in the unidentified charged particle analysis, 
the peripheral data are well described 
with an interpolation of existing hadron-hadron data\cite{Drees}
scaled by the number of binary collisions. 
(An additional 
factor derived from p-p data of 
$ \pi^0/(h^+ + h^-) = 1/3.2$ 
is used to scale the measured yield of unidentified
charged particles to neutral pions.) 
Once again, the central data
fall well below the corresponding  yield expected if 
high $p_T$ particle production scaled with the number of binary
collisions. This is strikingly illustrated by the scaled  ratio 
of Equation~\ref{Eq:ScaledRatio} for the
central to peripheral yields, as shown in Figure~\ref{Fig:Pi0Ratio}.

A comparison of the ratio in Figure~\ref{Fig:ChargedRatio} for unidentified charged
particles to that of Figure~\ref{Fig:Pi0Ratio} for $\pi^0$'s suggests that
the suppression may be more pronounced in the case of identified (neutral) pions.
Should this be established (the present understanding of the systematic errors 
does not permit
a definite conclusion) 
a consistent description would imply that
charged pions are also preferentially suppressed as compared to
the unidentified charged particles. 
Hints of precisely this behavior are seen 
in Figure~\ref{Fig:SPECTRAFig1.ps}, where the trends in identified particle
spectra suggest that the contributions from protons and anti-protons 
become comparable to those from pions for $p_T > 2$~GeV/c.
Further work is required to improve both statistical and especially
systematic errors to determine if the agreement of these data\cite{David}
with ``jet quenching'' predictions\cite{Wang} is in fact evidence
for an enhanced energy loss mechanism in hot nuclear matter.
Clearly essential to that program is a set of detailed measurements
of both p-p reference data and proton-nucleus collisions to determine
the quantitative value of the Cronin effect at RHIC energies.

\section{FUTURE PLANS}
\begin{figure}[hbt]
\vspace*{-11.0cm}
\begin{flushright}
\begin{minipage}[t]{0.4\textwidth}
\resizebox{1.0\textwidth}{!}{\includegraphics{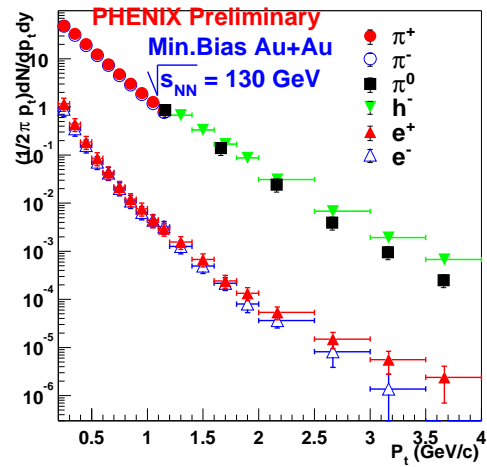}}
\caption{The inclusive spectrum of unidentified hadrons, $\pi^\pm$,
         $\pi^0$'s and $e^\pm$ measured in Au-Au minimum bias collisions.}
\label{Fig:Electrons}
\end{minipage}
\end{flushright}
\end{figure}

\begin{minipage}[b]{75mm}
A major component of the PHENIX physics program is dedicated to measurement
of leptonic signals. For example, the $\pi^0$ spectra described in 
Section~\ref{Sec:HighPT}, 
while intriguing and valuable in themselves, 
are also a prelude to the measurement of direct photons. 
In the central arms, virtual photons and vector mesons can be detected
via their decays to $e^+e^-$ pairs, while open charm and bottom production
are expected to dominate the production of single electrons for 
\hbox{$p_T > $2~GeV/c.} Measurement of these signals requires $\pi/e$ rejection
in excess of $10^3$ and very careful control of background contributions.
Figure~\ref{Fig:Electrons} shows the first efforts in this program\cite{Akiba}.
Charged particles are tracked using the PHENIX drift and pad chambers.
Electrons are selected by requiring that at least three PMT's fire in the RICH,
and are further identified via a tight matching cut between tracking momentum
and energy deposition in the EmCal. 
\end{minipage}

\begin{figure}[h]
 \centerline{
   \resizebox{0.43\textwidth}{!}{\includegraphics{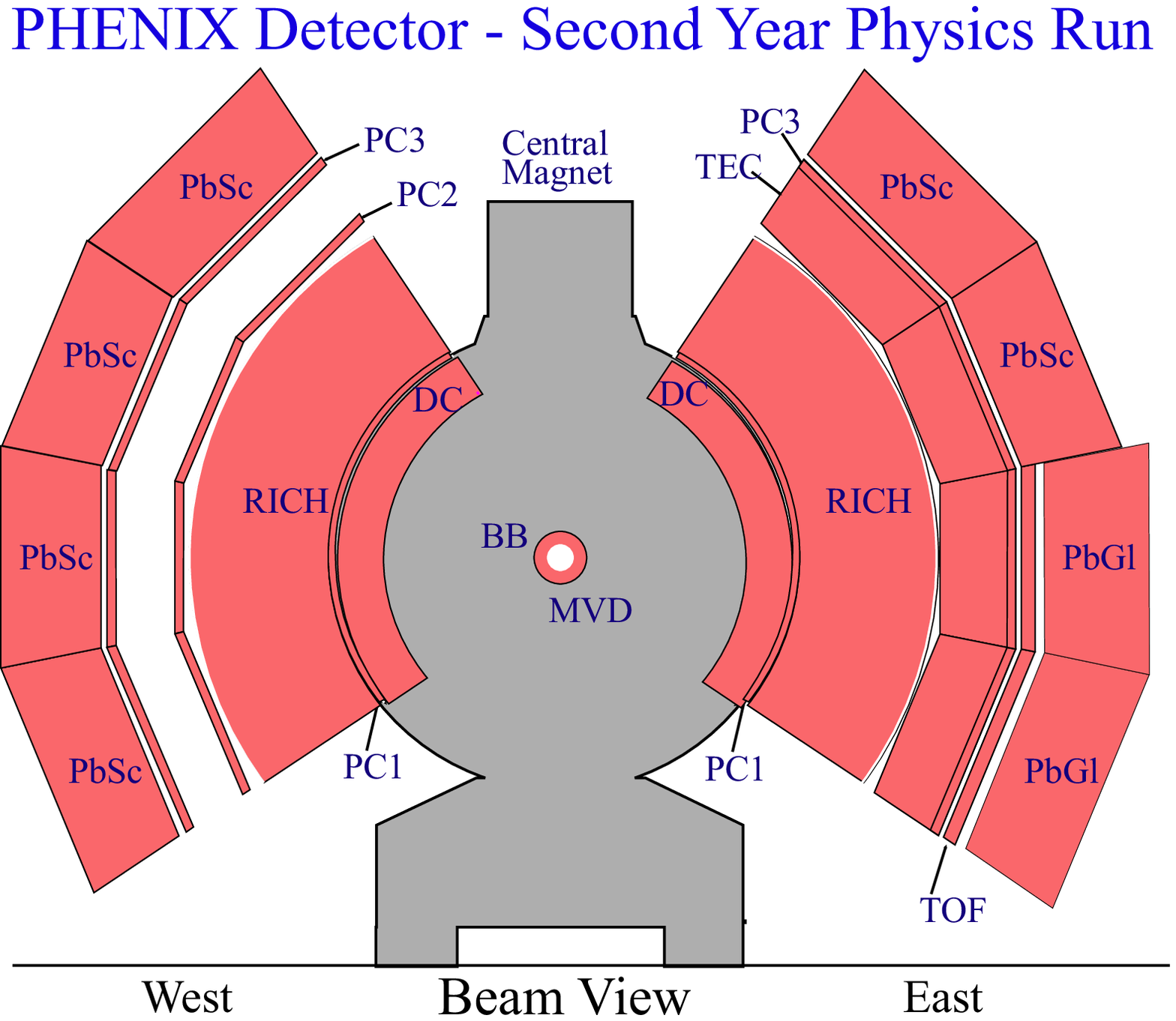}}
   \resizebox{0.57\textwidth}{!}{\includegraphics{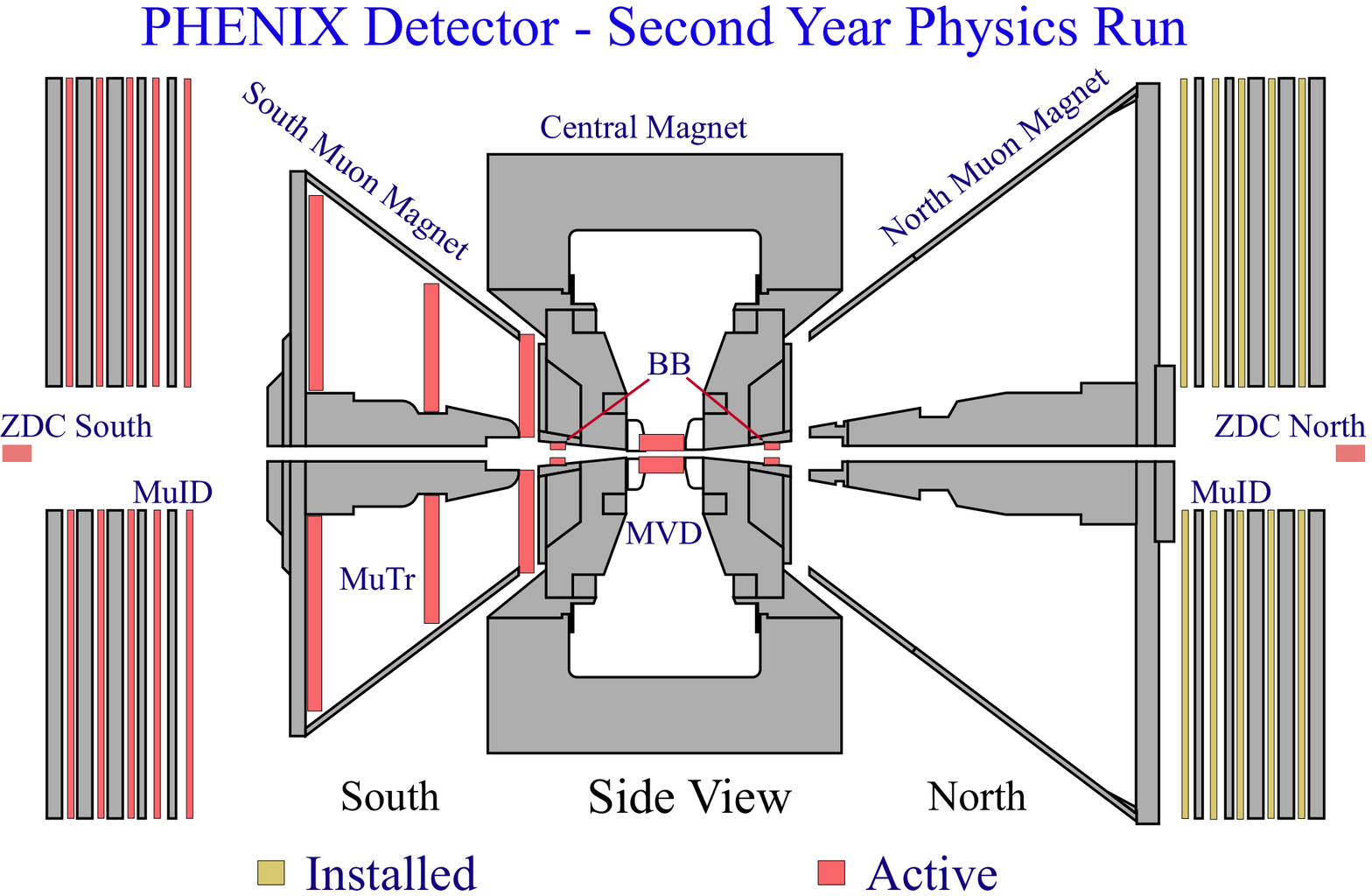}}
 }
  \caption{Installed and active detectors for  
           the RHIC Run-2 configuration of the PHENIX experiment.}
  \label{Fig:Run2}
\end{figure}

The power of this  combined particle identification from the various
PHENIX sub-systems is apparent from Figure~\ref{Fig:Electrons}, where
a clean electron spectrum is extracted at a level 2-3 orders of magnitude
below that of all charged particles.
This approach will form the basis for future PHENIX measurements of 
vector mesons and open charm at RHIC, and is also applicable to
photon measurements via external conversion.
The measurement of electrons along with all other PHENIX analyses 
will benefit greatly from the 
significantly enhancements made to the detector for Run-2 at RHIC.
As shown in Figure~\ref{Fig:Run2}, 
the complete aperture of the central arms will be available (compare to Figure~\ref{Fig:Run1}),
much of the MVD will be instrumented, and an entirely new spectrometer to measure muons
will be deployed. These additions in aperture and in capability, 
coupled with significant upgrades to the data
acquisition and triggering system, should result in a hundredfold or more increase
in the event sample obtained from Run-2. 
This will allow PHENIX to explore new signals and observables, to greatly
increase statistical precision and understanding of systematic errors on
existing analyses, to obtain the vital p-p comparison data, and to begin
a program of measurements with polarized protons dedicated to understanding
the proton spin.

\section{ACKNOWLEDGMENTS}
 
We thank the staff of the RHIC project, Collider-Accelerator, and Physics
Departments at BNL and the staff of PHENIX participating institutions for
their vital contributions.  We gratefully acknowledge support from the Department of
Energy and NSF (U.S.A.), Monbu-sho and STA (Japan), RAS, RMAE, and RMS
(Russia), BMBF and DAAD (Germany), FRN, NFR, and the Wallenberg Foundation
(Sweden), MIST and NSERC (Canada), CNPq and FAPESP (Brazil), IN2P3/CNRS
(France), DAE (India), KRF and KOSEF (Korea), and the US-Israel Binational
Science Foundation.

\end{document}